\documentclass[showpacs,showkeys,twocolumn,preprintnumbers,amsmath,amssymb,superscriptaddress]{revtex4-1} 
\usepackage[colorlinks, citecolor=blue]{hyperref}
\usepackage{graphicx}
\usepackage{subfigure} 
\usepackage{dcolumn}
\usepackage{bm}
\usepackage{color}
\usepackage{epstopdf}
\usepackage{float}
\usepackage{CJK}

\begin{document}
﻿
\begin{CJK*}{UTF8}{gbsn}
\preprint{Commun. Theor. Phys. \textbf{68}, 798 (2017)}

\title{Influence of cell-cell interactions on the population growth rate in a tumor}

\author{Yong Chen (陈勇)}
\affiliation{Center of Soft Matter Physics and its Applications, Beihang University, Beijing 100191, China}
\affiliation{School of Physics and Nuclear Energy Engineering, Beihang University, Beijing 100191, China}

\date{\today}

\begin{abstract}
The understanding of the macroscopic phenomenological models of the population growth at a microscopic level is important to predict the population behaviors emerged from the interactions between the individuals. In this work, we consider the influence of the population growth rate $R$ on the cell-cell interaction in a tumor system and show that, in most cases especially small proliferative probabilities, the regulative role of the interaction will be strengthened with the decline of the intrinsic proliferative probabilities. For the high replication rates of an individual and the cooperative interactions, the proliferative probability almost has no effect. We compute the dependences of $R$ on the interactions between the cells under the approximation of the nearest neighbor in the rim of an avascular tumor. Our results are helpful to qualitatively understand the influence of the interactions between the individuals on the growth rate in population systems.\\


\end{abstract}


\maketitle

\section{Introduction} 

The population growth is an important and widespread phenomenon in many areas of knowledge, such as economics, sociology, biology, etc. The growth curves are described by various empirical models by fitting the great variety of statistical data. For instance, the early Malthus model only depending on the number of individuals is of the exponential growth of the human population with small size~\cite{malthus1798}. With the modification with an exponentially decaying growth rate on the original Malthus's model, the asymmetric sigmoid growth curve of the Gompertz model describes the human lifespan and successfully apply to other biological systems~\cite{gompertz25,haybittle98,ausloos12}. The classical and well-fitting Verhulst model (somewhere named as the Logistic model), in many empirical data (bacterial growth, human population growth, etc.) leads to a logistic growth~\cite{verhulst38,verhulst45,verhulst47}. There are several other models known by the names of the authors, Ricker, Hassell, Beverton-Holt, Maynard-Smith, Richards, Bertalanffy and so on~\cite{boukal02,santos15}. 

All the above mentioned phenomenological models are based on the empirical data and only take into account the macroscopic information. It is well known that the macroscopic collective behaviors emerge from the interactions of the microscopic components of the studied systems. Unlike the simple interaction forms in science, the correlation between two living individuals is complicated and is hard to present a quantitative definition. Mombach et al. recapitulated some well-known models (the Malthus, Verhulst, Gompertz, and Richards models) in the viewpoint of starting from the microscopic properties under the assumptions of the fractal structure and the distance-modulation of the local interaction~\cite{mombach02}. After that, Ribeiro et al. presented a further analysis based on physical principle and developed the correspondence between the macroscopic growth rate and the microscopic distance function~\cite{santos15, ribeiro151,ribeiro152}. Obviously, this approximation of the interaction between individuals totally ignored the various characteristics of each individual.

Cell colony is a population of cells. Avascular tumor growth as a kind of cell growth is the initial stage of cancer development which is much simpler to construct mathematical models and to realize the quantitative experiments in higher reproducibility~\cite{roose07,tian10}. The tumor cell only has the following phenotypic states: proliferation, invasion, quiescence, and death. It means that the variety of the cell-cell interactions can be reduced by labeling the cell with its phenotype. Indeed the tumor cell is able to produce the cytotoxic substances against other cells and to influence the others by the intracellular chemical communication~\cite{tomlinson971,tomlinson972,kumar96,mansury06}. The simulation results based on the agent-based tumor model shown that the cell-cell interactions can influence both the population growth dynamics and the surface roughness~\cite{mansury06,chen15}. We believe that the phenotypic classification of the tumor cells is a reasonable and feasible way to model the population growth of a tumor at the level of single cells. 

Population growth rate as the basic crucial parameter in population dynamics is the rate at which the number of individuals in a population increases in a given time period. As far as we know, it is rare to study the influence of the microscopic individual-individual interaction on this parameter, probably because there is an important difficulty. In this work, we use a simple parameter to measure the effect of the cell-cell interaction on the intrinsic replication rate of the cells in a tumor and present the corresponding sequences of the growth rate. Then, in the case of the avascular tumor, we discuss the modulation of the growth rate by the intracellular phenotype-phenotype interaction.

\section{The model} 
In general, based on the population balance the number of cells of the type $i$ in a tumor evolves as the following continuous equation~\cite{martins07}，
\begin{eqnarray}
    \frac{\partial N_i(r,t)}{\partial t} &=& \mathcal{S}_i(r,\sigma,t) \nonumber \\
    &+& \mathcal{M}_i(r,\sigma,t) + \mathcal{P}_i(r,\sigma,t) + \mathcal{D}_i(r,\sigma,t) .
\label{eq01}
\end{eqnarray}
Here, $\mathcal{S}$ denotes the source/sink process, and $\mathcal{M}, \mathcal{P}$ and $\mathcal{D}$ are associated with invasive, proliferative and death phenotypic behaviors, respectively. 
All the terms on the right-hand side of the above equation are dependent on the location $r$ and the current functional state $\sigma$. 

Here, we only focus on the population size of a tumor, or the total amount of the tumor cell, $N(t) = \sum_i\int drN_i(r,t)$. For a short time interval $\Delta t$, the population of the tumor is updated with
\begin{equation}      
    N(t + \Delta t) = N(t) + \Delta N(t), 
    \label{eq02}
\end{equation}
where $\Delta N(t)$ is the changes in amount of cell during $\Delta t$. For simplicity, $\Delta N(t)$ depends on the cellular proliferation and death, $\mathcal{P}$ and $\mathcal{D}$, and is proportionate to the population size and the time interval. So, returning to the Eq.~(\ref{eq02}), we have 
\begin{eqnarray}      
    N(t + \Delta t) &=& N(t) + \Delta t RN(t).
    \label{eq03}
\end{eqnarray}
Here $R$ is the population growth rate. One takes the limit of infinitesimal $\Delta t$, converting the above discrete model (\ref{eq03}) to 
\begin{equation}      
    \frac{dN}{dt} = RN.     
    \label{eq04}
\end{equation}
Essentially, $R$ is not a simple constant. Based on fitting the statistical population data or considering the microscopical biological processes, several modified models have been proposed to describe the population dynamics. The widely used models are known as the Malthus, logistic, Gompertz, Richards-like, Ricker, Hassell model, and so on~\cite{boukal02}. All of the models basically are phenomenological descriptions in the macroscopic level. It is known that the collective macroscopic behavior emerges from the interaction of the microscopic components of the system. In this work, we apply the idea of emergent behavior to studying the population growth at the level of the individual cell, especially the contribution of the interaction among the cells to the growth rate.     

\section{Results and discussion} 
\subsection{Population growth rate and the intrinsic replication rate} 

For an individual cell, the intrinsic replication rate $R_0$ is regulated by a competition between the birth and the death (the tendency to proliferation or death). Mathematically, the intrinsic replication rate of a cell is 
\begin{equation}
    R_0 = p_0 - q_0,
    \label{eq05}
\end{equation}
where $p_0$ and $q_0$ are the proliferative  and the necrotic probabilities, respectively. Moreover, note that the probability normalization $p_0+q_0+m_0 =1$ and $m_0$ is the probability for others functional states of the cell, such as migration, invasion, quiescence, etc. 

\begin{figure*}[t]
    \centering
    \includegraphics[width=0.4\linewidth]{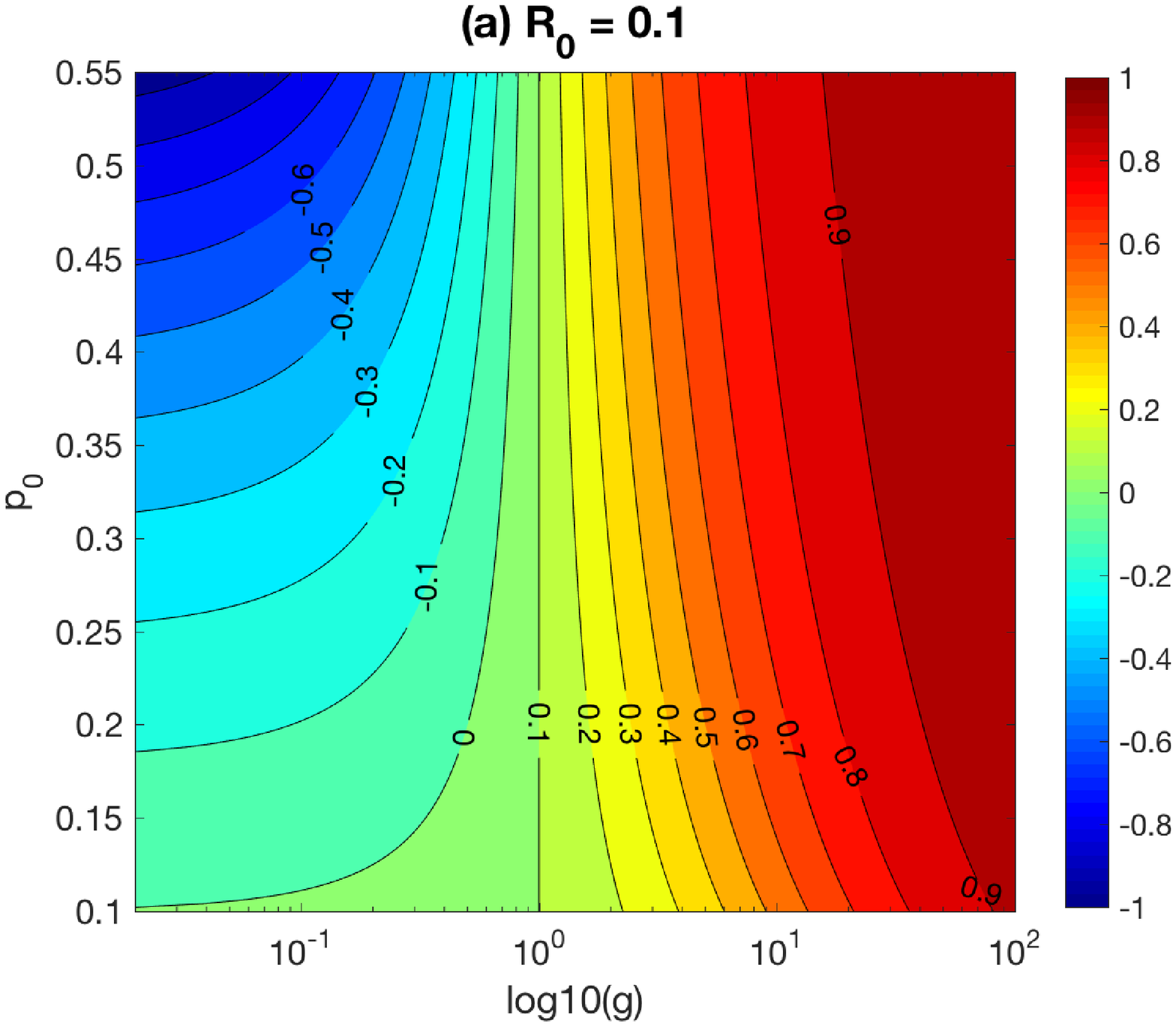}
    \includegraphics[width=0.4\linewidth]{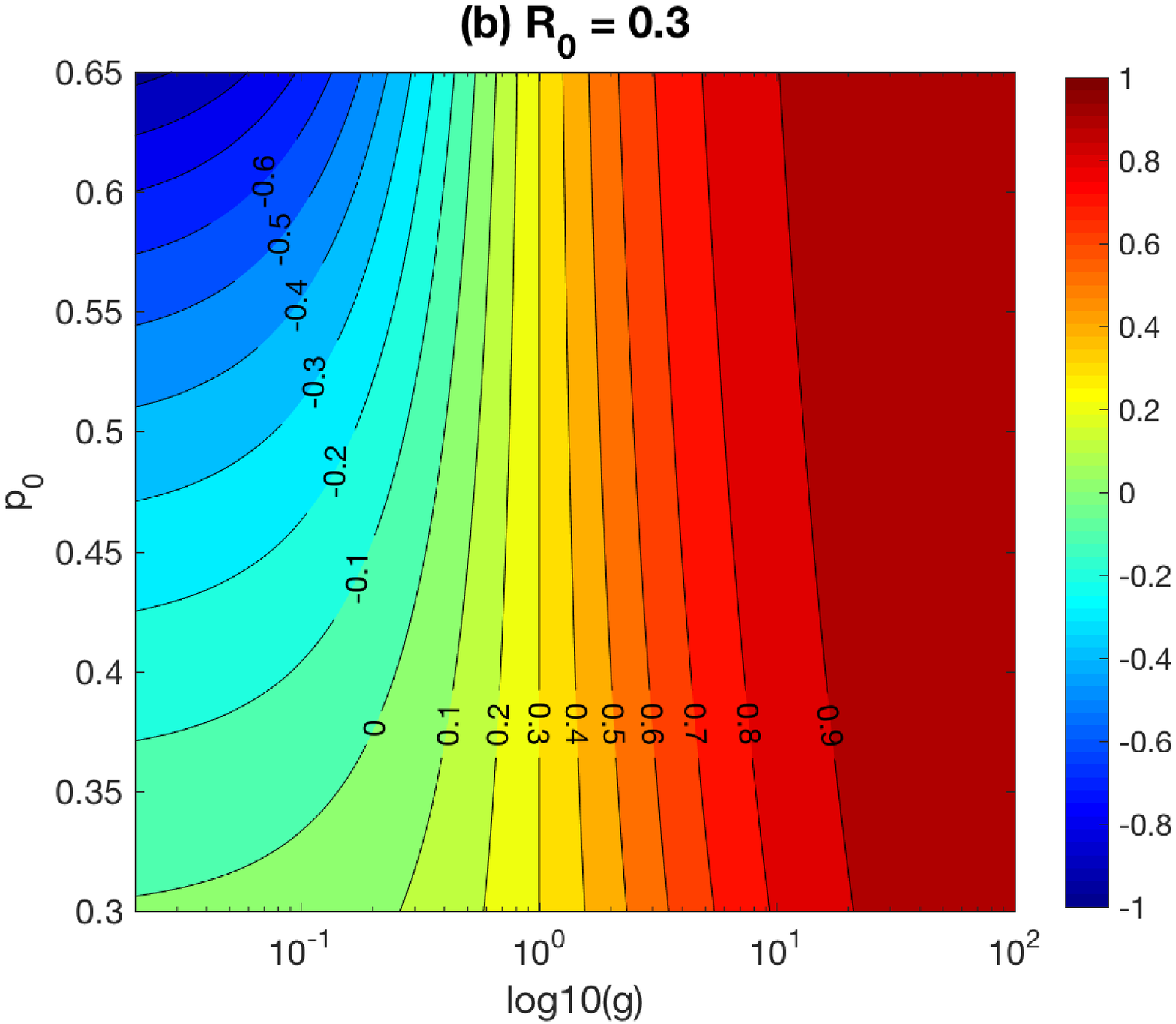}\\
    \includegraphics[width=0.4\linewidth]{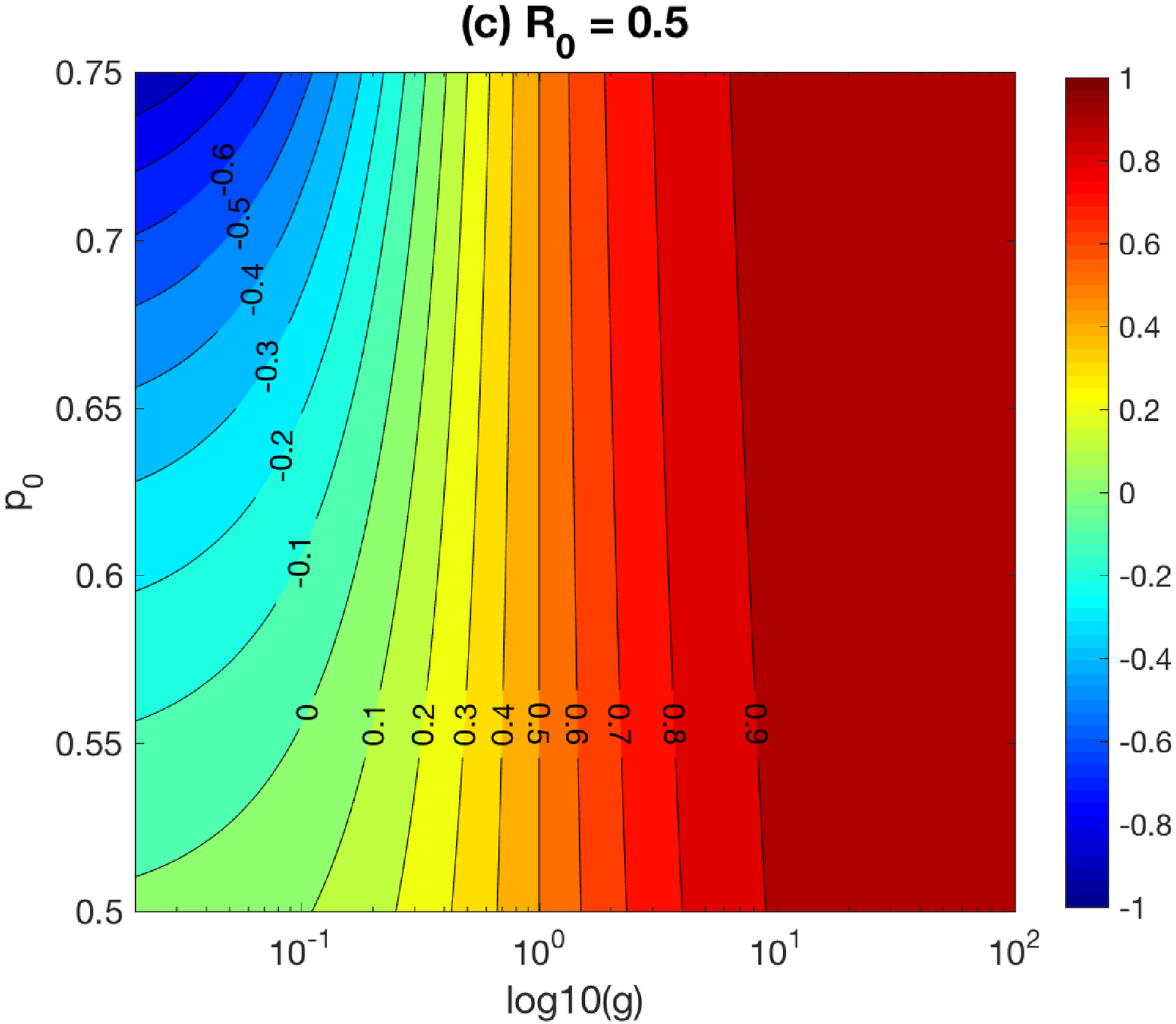}
    \includegraphics[width=0.4\linewidth]{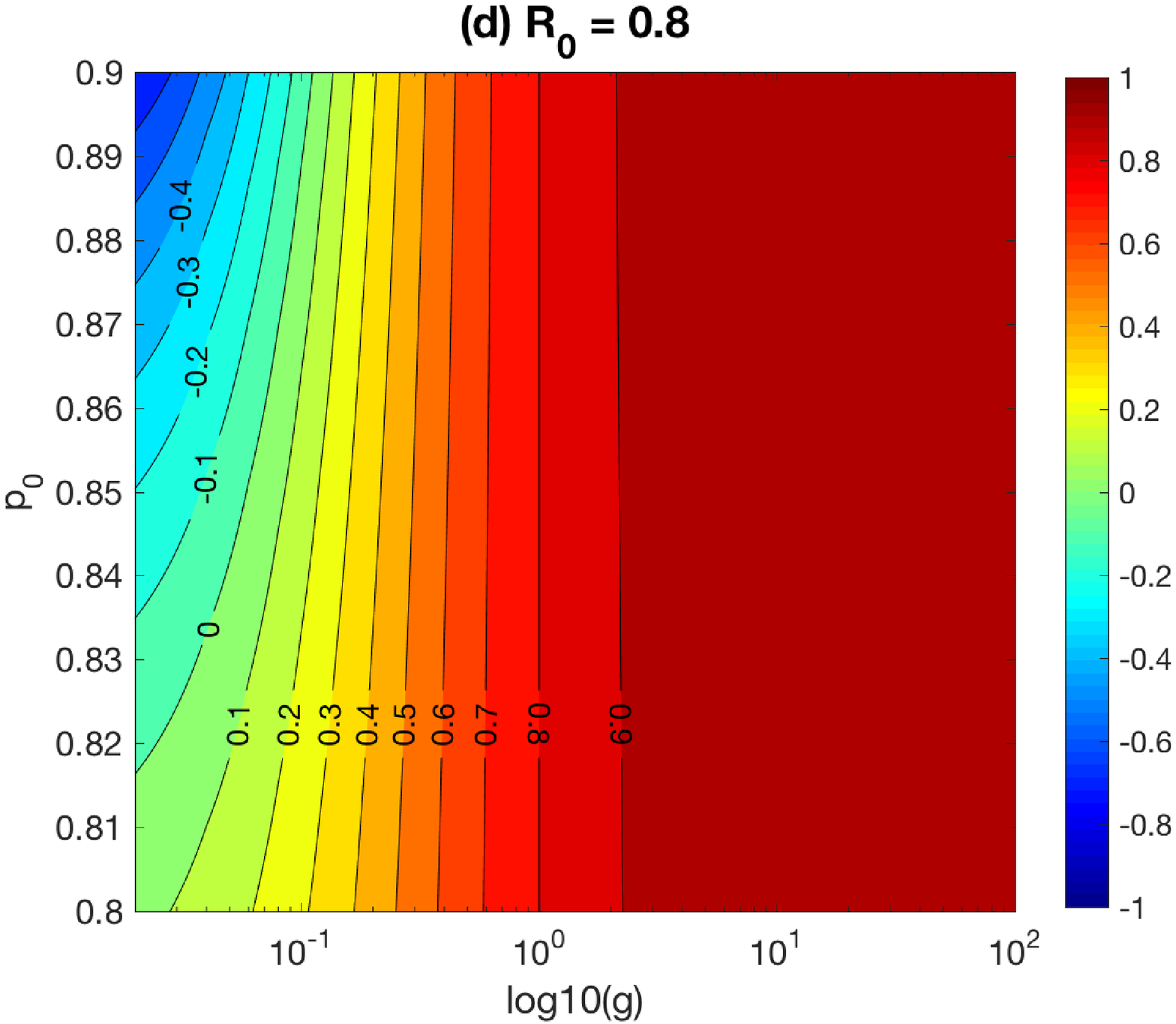}
    \caption{Contour graphs of the population growth rate $R$ involving the contribution from cell-cell interactions $g$ for different intrinsic growth rates (a) $R_0=0.1$, (b) $R_0=0.3$, (c) $R_0=0.5$, and (d) $R_0=0.8$. 
    }
    \label{fig1}
\end{figure*}

The update of the population obeys the following equation,
\begin{equation}
    \frac{dN}{dt} = \int dr R\rho_c(r,t) =\int dr R_0\rho_c(r,t),
    \label{eq06}
\end{equation} 
where $\rho_c(r,t)$ is the spatial distribution of the cells at time $t$. Note that $R=R_0$ in the above equation is only valid for the intrinsic replication of the cells. 

Clearly, the influence of the interaction among the cells on the population growth rate $R$ is the inevitable objective fact. For simplicity, we use $p=gp_0$ to describe the resulting proliferative probability which includes the intrinsic probability and the local environmental influences. This simplification is realistic since all the external influences finally appear as the intracellular biological processes. The parameter $g$ is larger than $0$. $g=1$ denotes the case without the contribution of the local field. $g<1$ means the interaction among the local cells is competitive, and $g>1$ is for the cooperation. Thus, the resulting replication rate of the individual cell of the population is      
\begin{eqnarray}
    R &=& \frac{gp_0 - q_0}{gp_0+q_0+m_0} = \frac{(g-1)p_0 + R_0}{(g-1)p_0+1}.
    \label{eq07}
\end{eqnarray}
It is obvious that the resulting rate $R$ is reduced to $R_0$ for $g=1$.

Figure~\ref{fig1} presents the dependences of the resulting replication rate $R$ on the intrinsic proliferative probability $p_0$ and the parameter describing the interactions among the cells $g$ under several intrinsic replication rates of the individual cell $R_0$. The range of $p_0$ depends on the value of $R_0$ and it follows $p_0 \in \left[R_0, (R_0+1)/{2} \right]$. Note that the horizontal axis is the data of $g$ in logarithmic coordinates. 

The lines of $g=1$ divide the phase diagrams into two parts, the left cooperation region, and the right competition region. In all cases of cooperation ($g>1$), the intrinsic proliferative probability $p_0$ has no obvious influence on the final replication rate $R$, especially in the region of larger $R_0$ and $p_0$. It only plays little role in the case of very small $R_0$ and $p_0$ (see Fig.~\ref{fig1}a). However, increasing $g$ is effective for improving the resulting $R$ and is efficient in cases of smaller $R_0$ and larger $p_0$. The left competition part in Figure~\ref{fig1} ($g<1$) is much different than the right one. First, there exists a contour line with $R=0$ which divides the left part into two regions, the bottom left region of population decrease ($R<0$) and the upper right region of population growth ($R>0$). As a result, it is possible that the tendency of population variation will switch over by changing $p_0$ or $g$. Interestingly, there is a minimum value of $p_0$ and the variation tendency cannot be changed if $p_0$ is smaller than it. For stronger competition from local field (smaller $g$), the role of $p_0$ regulating $R$ becomes remarkable. Similarly, $g$ is much more efficient in the cases of smaller $R_0$ and larger $p_0$. Comparatively speaking, in most cases (in the vicinity of $g=1$), changing $g$ is a more feasible and efficient solution to regulate the resulting replication rate $R$.    

\subsection{Population growth rate in an avascular tumor} 
All tumor progression undergo a stage of avascular growth. The main feature of avascular tumors is composed of a necrotic core of dead tumor cells due to nutrient starvation and an outer proliferative rim with abundant nutrient environment~\cite{martins07,roose07}. In another word, there is almost no necrotic cell in the outer rim of the avascular tumor. According to the definition of the replication rate in Eq.~(\ref{eq05}), the population size $N$ in equation~(\ref{eq06}) is the total number of the alive cells. Thus, it is feasible to assume $q_0 \simeq 0$ and $R_0 \simeq p_0$ since the growth rate of avascular tumor is determined by the growth of non-necrotic outer rim. One can rewrite the resulting replication rate in expression~(\ref{eq07}),
\begin{equation}
    R \simeq \frac{gp_0}{(g-1)p_0+1} = \frac{gR_0}{(g-1)R_0+1}.
    \label{eq08}
\end{equation}
The equation above presents there is no dead cells in the local interaction field.

\begin{figure*}[t]
    \centering
    \includegraphics[width=0.4\linewidth]{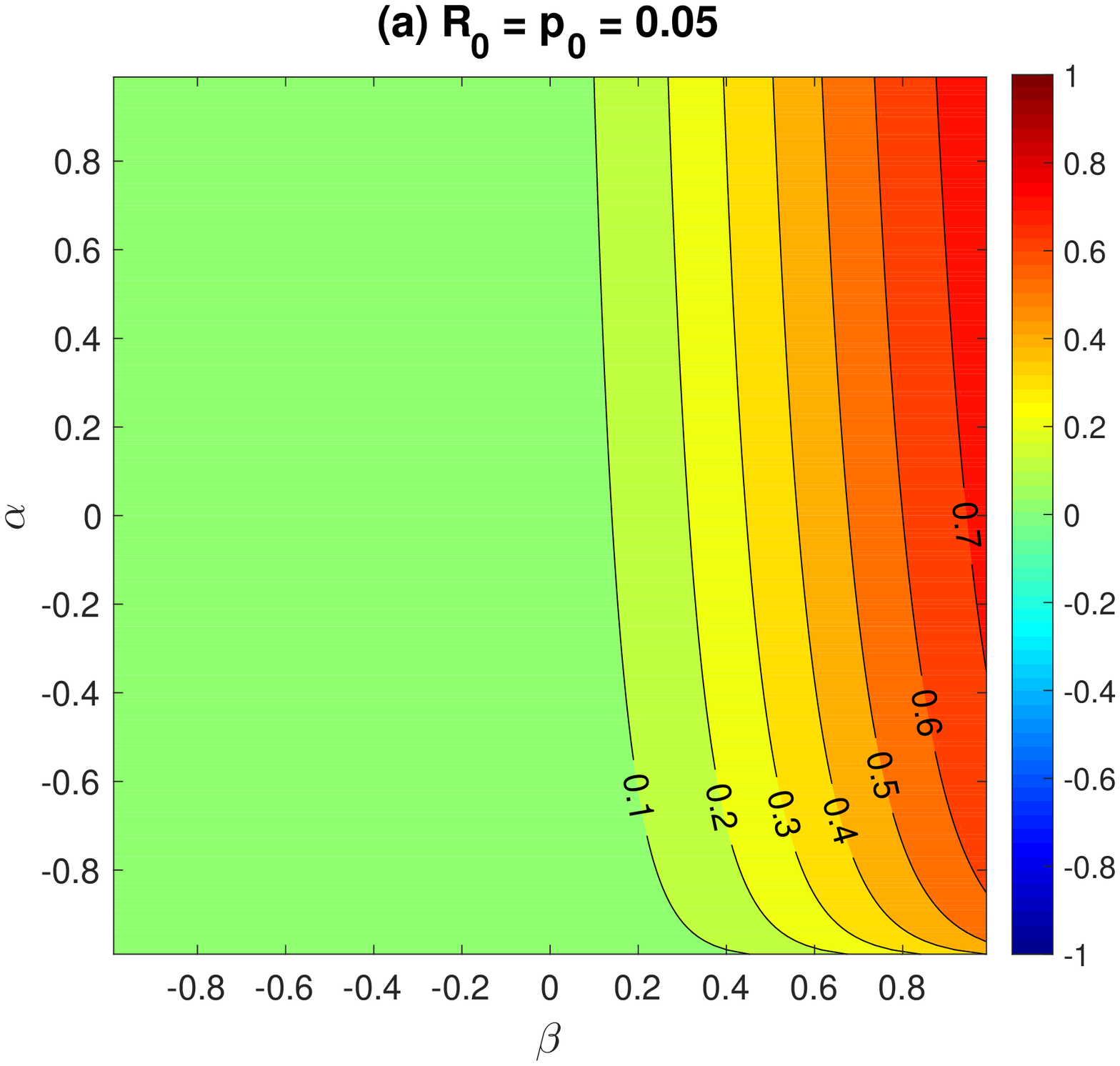}
    \includegraphics[width=0.4\linewidth]{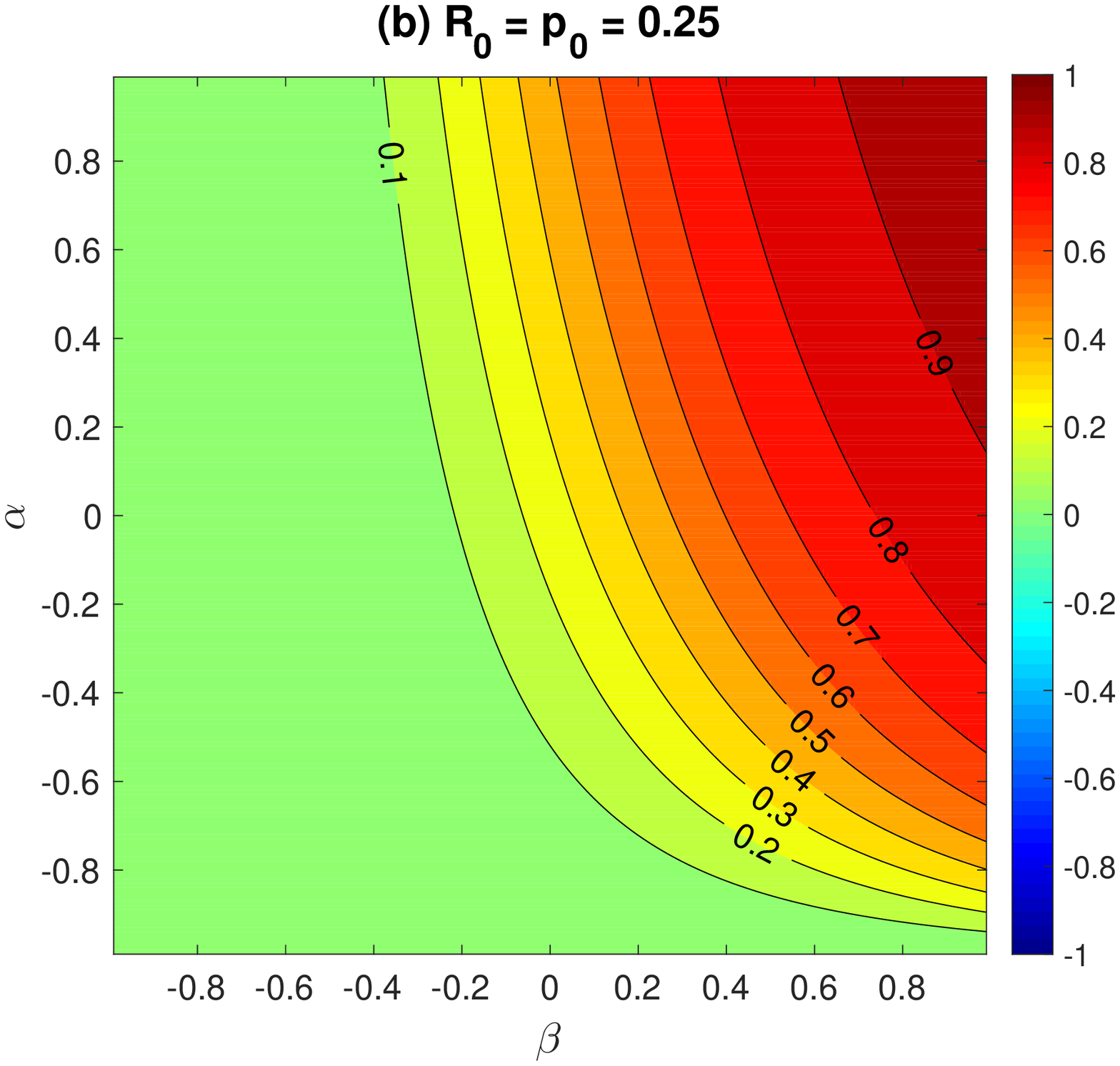}\\
    \includegraphics[width=0.4\linewidth]{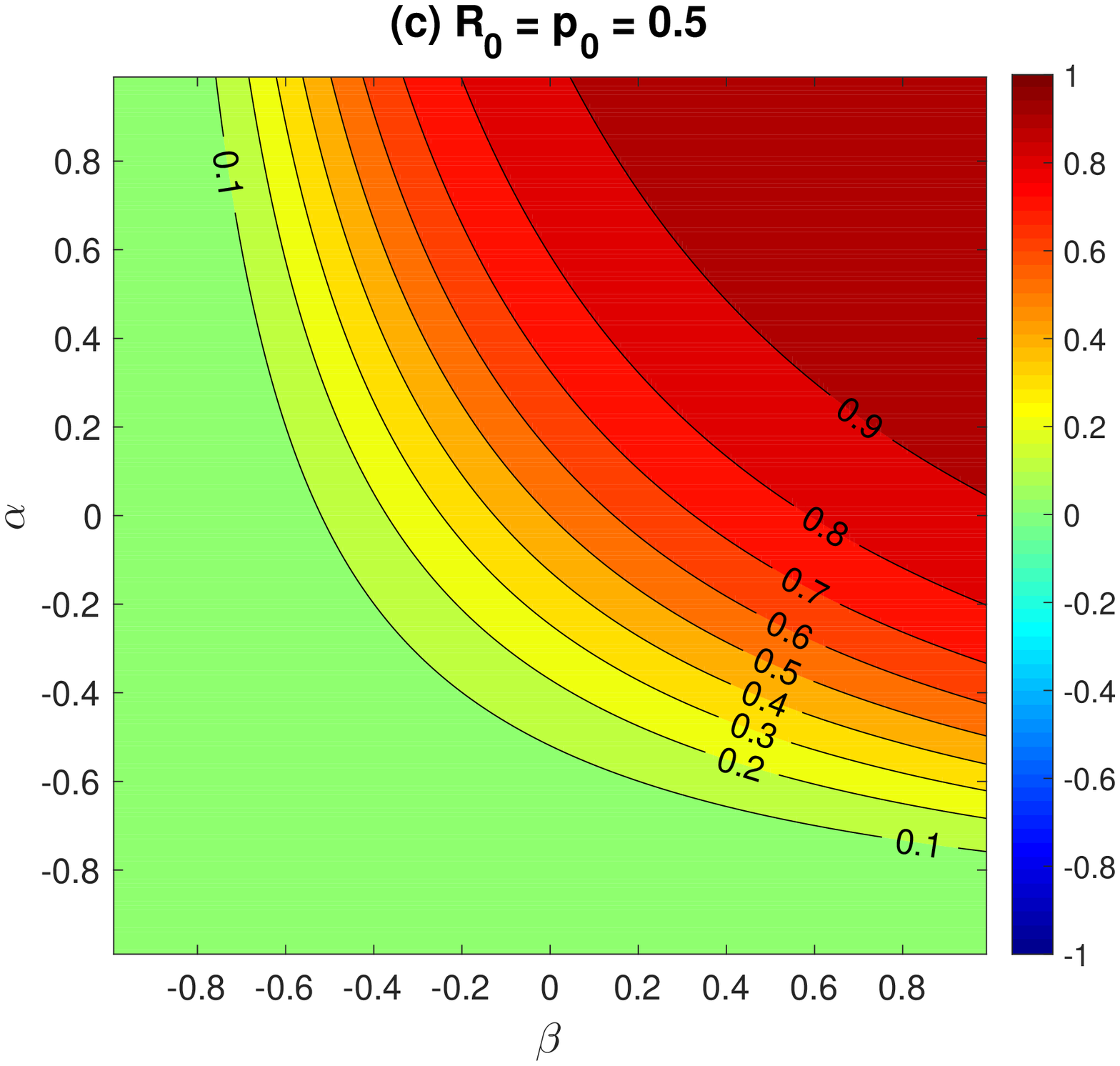}
    \includegraphics[width=0.4\linewidth]{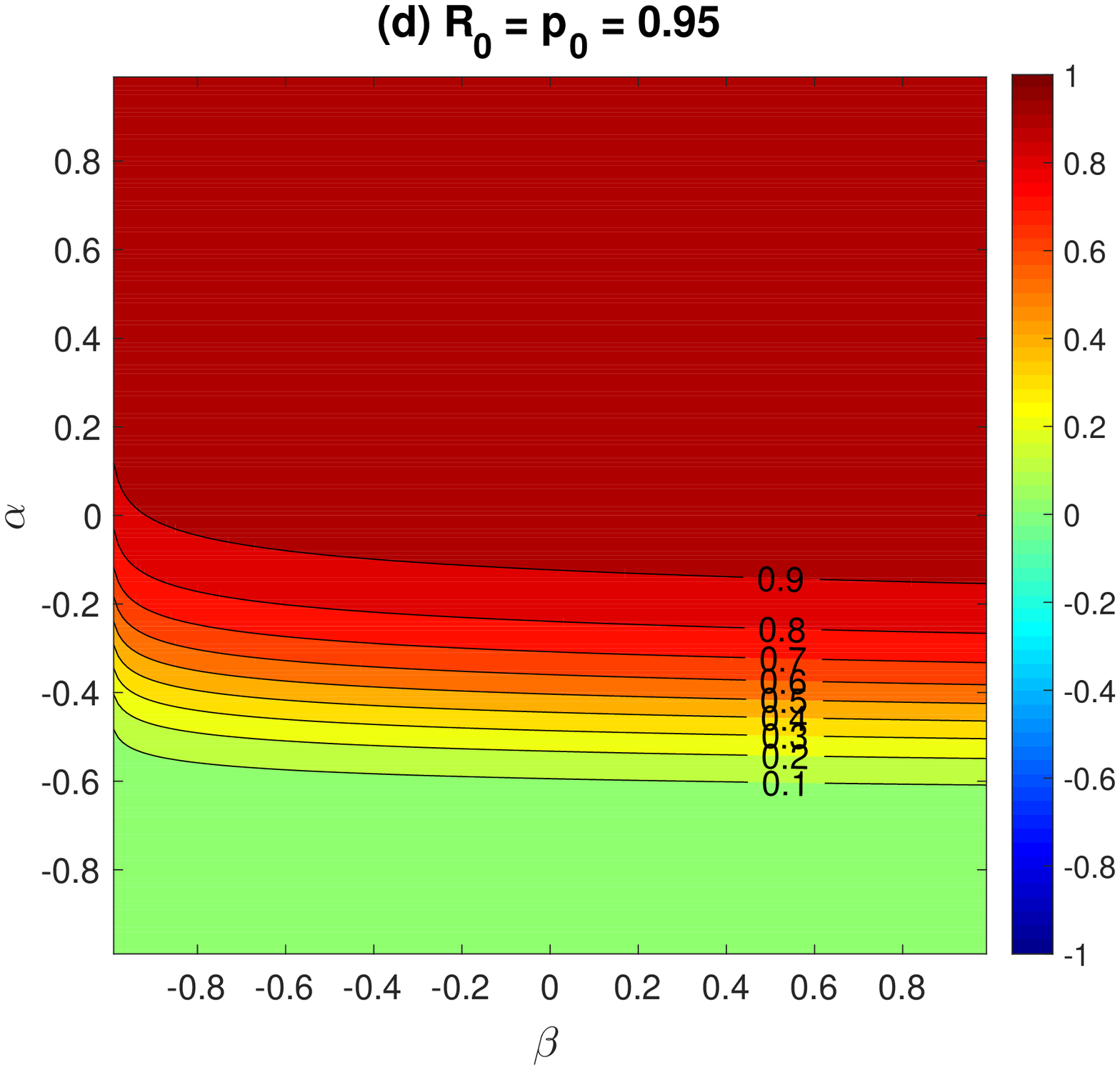}
    \caption{Contour graphs of the population growth rate $R$ involving the contributions of the combination of two kinds of cell-cell interactions $\alpha$ and $\beta$ for different intrinsic growth rates (a) $R_0=0.05$, (b) $R_0=0.2$, (c) $R_0=0.5$, and (d) $R_0=0.95$. Here, we choose $N_0 =6$ for two dimensional avascular tumors.   
    }
    \label{fig2}
\end{figure*}

The interaction parameter $g$ obviously is associated with a number of factors, such as the local cell number density, the nutrient concentration, the cellular functional state, and so on. Inspired by the packing structure in solid crystal, we consider the contribution of the nearest neighbor cells on the local interaction. The interaction parameter will have the form~\cite{mansury06,chen15}   
\begin{equation}
    g = \prod_i^{p_0 N_0} (1+\alpha_i) \prod_j^{(1-p_0)N_0} (1+\beta_j). 
    \label{eq09}
\end{equation}
Here, $\alpha_i \in (-1, 1)$ denotes the interaction between two proliferative cells and $\beta_j \in (-1, 1)$ describes the interaction between a proliferative cell and another cell with other phenotypes. Note that the mutual interaction between two cells normally brings into effect by the autocrine/paracrine growth factor, the cytotoxic substance, the chemical communication, etc. and the linking genotypes are complex. $N_0$ presents the total number of the nearest neighbor cells. For example, $N_0=6$ for two dimensional triangel structure and $12$ is for three dimensional hexagonal close-packed (HCP) scenario. Furthermore, neglecting the cell type diversity and assuming that $\alpha_i =\alpha$ and $\beta_j = \beta$, the interaction parameter in Eq.~(\ref{eq09}) can be approximated by 
\begin{equation}
    g \simeq  (1+\alpha)^{p_0 N_0}(1+\beta)^{(1-p_0)N_0}.
    \label{eq10}
\end{equation}

Substituting the expression $g$ above to Eq.~(\ref{eq08}), we obtain the dependence of the growth rate $R$ on the different cell-cell interactions $\alpha$ and $\beta$ as shown in Figure~\ref{fig2}. In all cases, $R>0$ means a population growth process since the approximation the death probability $q_0=0$ in the proliferative rim of the avascular tumor. Both $\alpha >0$ and $\beta >0$ denote the cooperative interactions. 

Actually, based on the equation~(\ref{eq09}), one can find changing $\alpha$ or $\beta$ will make a similar influence on $R$ which is shown in Fig.~\ref{fig2}(c). The stronger the positive/negative $\alpha$ ($\beta$), the larger/smaller $R$. In the extreme case of very low intrinsic replication rate ($R_0$ is very small illustrated in Figure~\ref{fig2}a), only the interaction between the proliferative cell and the cell at other functional states $\beta$ has a remarkable cooperative influence on $R$ because of the much smaller amount of the proliferative cell. In another extreme case of $R$ close to $1$ (see Fig.~\ref{fig2}d), the interaction between two proliferative cells $\alpha$ is effective to suppress $R$ because it has plenty of proliferative cells. 

It is worth to note that $N_0 =6$ and Figure~\ref{fig2} shows the results for two-dimensional scenario. We also observe the similar dependence of $R$ on $\alpha$ and $\beta$ in three-dimensional systems (HCP approximation and $N_0=12$). Typically, considering the fractal structure experimentally observed in tumors and in the growth of bacteria, the local cell amount is given by $N_0 = \rho_0 \Omega_D \int_0^{r_0} r^{D_F -1}dr $~\cite{cross97,fujikawa89}. Here, $\rho_0$ is a constant. $\Omega_D$ is the solid angle and $D_F$ is the fractal parameter. For fixed cut-off $r_0$, $N_0$ still represents a constant amount of the local cells involving the interactions. In addition, if taking account of the necrotic probability $q_0 \neq 0$, it is observed the similar tendencies in Figure~\ref{fig2} but the population reduction (the negative $R$) will emerge in the bottom-left parameter area.

\section{Conclusion and remarks}
In this work, we study the influence of the cell-cell interactions on the population growth rate in the level of the individual cell. By integrating the interaction from the surrounding of the cells into the changing of the intrinsic replication rate, we present the dependence of the resulting replication rate on the proliferative probability and the cell-cell interactions. It is found that the influence of the local field is remarkable for large intrinsic replication rate and proliferative probability. Moreover, in most cases changing the interaction between the cells is a more feasible recipe to regulate the population growth rate. Under the nearest neighbor approximation in the outer rim of two-dimensional avascular tumors, we divide the cell-cell interactions into two classes and conjecture that the local cell number density is crucial to influence the resulting replication rate.

Rigorously, the population growth rate $R$ in Eq.~(\ref{eq06}) is the replication rate of an individual in a population and is not the exact equivalent of the per-capita population growth rate $R$ in Eq.~(\ref{eq04}). The distribution of the cells $\rho_c(r,t)$ can be described by the fractal structure, or be solved by the continuous descriptions, such as the reaction-diffusion model, two-phase model, and so on~\cite{ferreira03,roose07,martins07}. By using equations~(\ref{eq07},\ref{eq09}) and solving equation~(\ref{eq06}), one will obtain the population growth equation and the per-capita growth rate.   

It is should be noted that the cell-cell interactions in present work can be used to study the other similar systems with a lot of individuals. The results presented in this work will cast some light on explaining the collective phenomena induced by the individual–individual interaction properties in a more fundamental way.

\section*{Acknowledgments}
This work was supported by the National Natural Science Foundation of China under Grant No. 11675008, and Grant No. 21434001.

\end{CJK*}

\end{document}